\documentstyle[axodraw]{article}

\def\beq{\begin{equation}}
\def\eeq{\end{equation}}

\def\gtap{\raisebox{-.4ex}{\rlap{$\sim$}} \raisebox{.4ex}{$>$}}
\newcommand{\Rsl}{{\not \! \!{R}}}

\newcommand{\ms}{\widetilde{m}}

\newcommand{\nuH}{\nu_{_H}}

\begin{document}
\vspace*{-1in}
\renewcommand{\thefootnote}{\fnsymbol{footnote}}
\begin{flushright}
SINP/TNP/00-7\\
CUPP-00/2 \\
\texttt{hep-ph/0004108} 
\end{flushright}
\vskip 5pt
\begin{center}
{\Large {\bf Cosmological constraints on $R$-parity violation \\ from
neutrino decay}}
\vskip 25pt
{\sf Gautam Bhattacharyya $^{a,\!\!}$
\footnote{E-mail address: gb@tnp.saha.ernet.in}},   
{\sf Subhendu Rakshit $^{b,\!\!}$
\footnote{E-mail address: srakshit@cucc.ernet.in}}, 
and 
{\sf Amitava Raychaudhuri $^{b,\!\!}$
\footnote{E-mail address: amitava@cubmb.ernet.in}}  
\vskip 10pt  
$^a${\it Saha Institute of Nuclear Physics, 1/AF Bidhan Nagar, 
Calcutta 700064, India}\\
$^b${\it Department of Physics, University of Calcutta, 92
Acharya Prafulla Chandra Road, \\ Calcutta 700009, India} 
\vskip 20pt
{\bf Abstract}
\end{center}

\begin{quotation}
{\noindent\small 
If the neutrino mass is non-zero, as hinted by several
experiments, then $R$-parity-violating supersymmetric Yukawa
couplings can drive a heavy neutrino decay into lighter states.
The heavy neutrino may either decay radiatively into a lighter
neutrino, or it may decay into three light neutrinos through a
$Z$-mediated penguin.  For a given mass of the decaying neutrino,
we calculate its lifetime for the various modes, each mode
requiring certain pairs of $R$-parity-violating couplings be
non-zero.  We then check whether the calculated lifetimes fall in
zones allowed or excluded by cosmological requirements. For the
latter case, we derive stringent new constraints on the
corresponding products of $R$-parity-violating couplings for
given values of the decaying neutrino mass.

\vskip 10pt
\noindent
PACS number(s): 13.35.Hb, 12.60.Jv, 11.30.Fs}

\end{quotation}

\vskip 20pt  

\setcounter{footnote}{0}
\renewcommand{\thefootnote}{\arabic{footnote}}

\section{Introduction}

The Super-Kamiokande collaboration has recently provided compelling
evidence in support of neutrino oscillations as an explanation of the
atmospheric anomaly \cite{superk}. The observed solar neutrino
deficit~\cite{solar} and the LSND accelerator experiment results
\cite{lsnd} are also indicative of oscillations. These results achieve
especial significance since oscillations require the neutrinos to be
massive. A massive neutrino signals physics beyond the standard model
and has far-reaching implications in particle physics, astrophysics,
and cosmology \cite{numass}.

In this work, we consider $R$-parity-violating ($\Rsl$)
supersymmetry (defined later) in the context of neutrino decays
\cite{emr,rt}. Such interactions violate lepton number and in the
presence of appropriate couplings of this type a heavier neutrino
of one flavour can decay to a lighter one of a different flavour
in association with the emission of a photon. Alternatively, the
heavier neutrino can decay invisibly into three lighter neutrinos
through one loop graphs involving $\Rsl$ couplings.  

Cosmological and astrophysical requirements forbid certain regions in
the neutrino mass and lifetime plane. They originate, for
example, (i) from the precise black-body nature of the microwave
background radiation, (ii) from the tight requirements of
consistency of the predicted primordial nucleosynthesis with
observation, etc., which are discussed later.

We calculate, for a heavy neutrino of a given mass, the decay
lifetime induced by $\Rsl$ couplings. If the lifetime so obtained
falls in the forbidden region then couplings of the chosen
strength are not allowed for the neutrino mass used. In this way
we can establish new constraints on some lepton number violating
couplings applicable for specific neutrino masses. On the other
hand, if one assumes that these $\Rsl$ couplings are at their
existing limits, then upper bounds can be set on the decaying
neutrino mass.

Before we move to the next section, a short introduction to $R$-parity
and $\Rsl$ supersymmetry \cite{rpar} is in order.  `$R$-parity' in
supersymmetry refers to a discrete symmetry which follows from the
conservation of lepton-number ($L$) and baryon-number ($B$). For
the standard model particles and their superpartners, it is
defined as $R=(-1)^{(3B+L+2S)}$, where $S$ is the intrinsic spin of
the field. $R$ is $+1$ for all standard model particles and $-1$ for
all superparticles.

The most general $\Rsl$ superpotential is given by,
\begin{equation}
 W_{\not \! {R}} 
={1 \over 2} \lambda_{ijk} L_{i} L_{j} E^{c}_{k}
           + \lambda'_{ijk} L_{i} Q_{j} D^{c}_{k}
          + {1 \over 2} \lambda''_{ijk} U^{c}_{i} D^{c}_{j} D^{c}_{k}
           +\mu_{i} L_{i} H_u, 
\label{superpot}
\end{equation}
where $i, j, k = 1, 2, 3$ are quark and lepton generation indices;
$L_i$ and $Q_i$ are $SU(2)$-doublet lepton and quark superfields
respectively; $E^{c}_i$, $U^{c}_i$, $D^{c}_i$ are $SU(2)$-singlet
charged lepton, up- and down-type quark superfields respectively;
$H_u$ is the Higgs superfield responsible for the generation of
up-type quark masses; $\lambda_{ijk}$ and $\lambda'_{ijk}$ are
$L$-violating while $\lambda''_{ijk}$ are $B$-violating Yukawa
couplings. $B$- and $L$- conservation are not ensured by gauge
invariance and hence there is {\em a priori} no reason to set these
couplings to zero.  $\lambda_{ijk}$ is antisymmetric under the
interchange of the first two generation indices, while
$\lambda''_{ijk}$ is antisymmetric under the interchange of the last
two. Thus there could be 27 $\lambda'$, 9 each of $\lambda$ and
$\lambda''$ couplings and 3 $\mu_i$ parameters. We assume that the
generation indices correspond to the flavour basis of fermions.  We
also note at this point that the $B$-violating couplings
$\lambda''_{ijk}$ and the bilinear couplings $\mu_i$ are not of any
relevance to our present analysis. We shall deal only with the
$L$-violating trilinear couplings and, for the sake of simplicity, we
assume that they are real.

Stringent constraints on individual $L$-violating couplings have been
placed from the consideration of neutrinoless double beta decay,
$\nu_e$-Majorana mass, charged-current universality, $e-\mu-\tau$
universality, $\nu_{\mu}$ deep-inelastic scattering, atomic parity
violation, $\tau$ decays, $D$ and $K$ decays, $Z$ decays, etc. Product
couplings (two at a time), on the other hand, have been constrained by
considering $\mu-e$ conversion, $\mu\rightarrow e\gamma$,
$b\rightarrow s \gamma$, $B$ decays into two charged leptons,
$K_L-K_S$ and $B_q-\overline{B_q}$ ($q = d,s$) mass differences, etc. (For
a collection of all these limits, see \cite{review}).

In the following section we elaborate on the neutrino mass
spectra which are preferred by the experimental data. In section
3 we discuss the radiative neutrino decay modes driven by the
$\lambda'$- and $\lambda$-type $\Rsl$ couplings while in section
4 we turn to the invisible three neutrino decay channel.  In the
next section we check if and how cosmological requirements on
these decay lifetimes constrain the $\Rsl$ couplings. We end in
section 6 with our conclusions.

\section{Neutrino mass spectrum}

In this section we elaborate on the neutrino masses that we will be
using to examine possible neutrino decays driven by $\Rsl$
interactions. We are motivated in making these choices by the
available evidence for massive neutrinos.

The upper limits on neutrino masses obtained from laboratory
experiments are the following. The masses of $\nu_{\mu}$ and
$\nu_{\tau}$ are constrained to be $m_{\nu_{\mu}} \leq 0.17$ MeV
and $m_{\nu_{\tau}} \leq 18.2$ MeV \cite{pdg}. As regards
$\nu_e$, the upper limit from tritium beta-decay end-point
measurements is 2.5 eV \cite{etrit}, while if it is of Majorana
nature then from the absence of neutrinoless double beta decay
one has the constraint\footnote{Strictly speaking, this bound
applies to the $(ee)$ element of the Majorana mass matrix written
in the flavor basis.} $m_{\nu_e} \leq 0.2$ eV
\cite{ebeta}.

Before we discuss the constraints that emerge from the
oscillation data, a few remarks are in order. The $\Rsl$
couplings have been defined in terms of flavour eigenstates. On
the other hand, any discussion of decay must refer to a neutrino
in its mass eigenstate. For simplicity of presentation and the
ease of illustration, in the following we do not distinguish
between the flavour and the mass eigenstates.  Indeed, the
indications for neutrino oscillation imply that this is not the
actual situation. In fact, the mismatch between the mass and
flavour bases is an essential ingredient of neutrino oscillations
and even maximal mixing is sometimes preferred (e.g., as required
by the atmospheric anomaly). It is straightforward to incorporate
the effect of this mixing in our results. One must multiply the
decay rates for $\nu_i \rightarrow \nu_{i'}$, presented later, by
appropriate factors determined by the probability of the $\nu_i$
($\nu_{i'}$) being present in the parent (daughter) neutrino.

In much of the analysis we choose the decaying neutrino to be
$\nu_\tau$ and for the purpose of illustration take its mass to range
from 45 eV to 100 keV. This is consistent with the laboratory bound
on $m_{\nu_\tau}$ \cite{pdg}. We consider its decay to either
$\nu_\mu$ or $\nu_e$. We also consider the
possibility of $\nu_\mu$ decaying to $\nu_e$. There are models in
the literature proposing {\em inverted} mass hierarchy scenarios
$m_{\nu_\tau} < m_{\nu_\mu} < m_{\nu_e}$. But, as far as bounds on the
$\Rsl$-couplings are concerned, these provide no new constraints since
the rate for $\nu_i \rightarrow \nu_{i'}$ decay ($i,i'$ flavour
indices) in this scenario is the same as that for $\nu_{i'}
\rightarrow \nu_i$ if the masses $m_{\nu_i}$ and $m_{\nu_{i'}}$ are
interchanged.

Now let us recollect what is known about the magnitude of
neutrino mass splittings from the data on neutrino oscillations.
There are two independent recent results -- namely, the
atmospheric neutrino anomaly \cite{superk} and the solar neutrino
problem \cite{solar} -- which can be conveniently explained
within the framework of neutrino oscillations. The LSND
experiment has also claimed positive evidence of $\bar{\nu}_\mu -
\bar{\nu}_e$ (and also $\nu_\mu - \nu_e$) oscillations
\cite{lsnd}.  However, this result is still awaiting independent
confirmation.

Oscillations require a non-zero mass splitting between neutrinos of
different flavour. More precisely, the experimental data determine
$\Delta m^2 = m_1^2 - m_2^2$, where $m_{1,2}$ are the masses of two
neutrinos. Oscillations also determine the unitary transformation
which relates the mass basis of neutrinos to the flavour basis. As
mentioned earlier, we are not directly concerned with this aspect and
do not discuss it any further\footnote{In this work we motivate
possible neutrino mass spectra from the experimental results and do
not attempt to reproduce the mass matrix structures that would fit the
data. There is a vast and growing literature in this subject related
to $\Rsl$ supersymmetry \cite{pap1}.}.

The information from experimental results on neutrino mass
splittings are as follows:

\begin{enumerate}
\item The atmospheric neutrino data can be explained in terms of
  oscillations of $\nu_\mu$ to either the $\nu_\tau$ or a sterile
  state $\nu_s$. The sterile neutrino does not couple to any of the
  standard model particles and passes undetected through the
  experimental set up. The necessary mass splitting has been found to
  be $\Delta m^2 \sim 10^{-2} - 10^{-3}$ eV$^2 \equiv \Delta_{\rm
    atmos}$.

\item The solar neutrino problem can be explained in terms of vacuum
neutrino oscillations of the $\nu_e$ to either the $\nu_{\mu,\tau}$ or
the $\nu_s$ with $\Delta m^2 \sim 10^{-10}$ eV$^2$. The
data admits an alternative solution in terms of MSW resonant neutrino
conversion if $\Delta m^2 \sim 10^{-5} - 10^{-6}$ eV$^2$. In the
latter case the sign of the mass splitting is fixed and the $\nu_e$
must be the lighter state of the two. We indicate the mass
splitting required for a solution to the solar neutrino problem by
$\Delta_{\rm solar}$.

\item The LSND experiment indicates $\nu_\mu - \nu_e$ oscillation
with $\Delta m^2 \sim 1$ eV$^2 \equiv \Delta_{\rm LSND}$.
\end{enumerate}

As is seen from the above, if both the atmospheric neutrino
anomaly and the solar neutrino deficit are sought to be explained
by invoking oscillations between three sequential neutrinos only,
then the mass splittings among them must be small.  On the other
hand, if one admits a fourth (sterile) neutrino in the model,
then it is possible to arrange for larger mass splittings. With a
fouth neutrino, one can have two categories of mass hierarchies.
In category `A', there are three closely spaced almost degenerate
states explaining the solar and atmospheric data and the fourth
one is separated by a larger gap $\Delta$. The fourth neutrino
could either be the sterile or $\nu_\tau$ since the $\nu_e$
($\nu_\mu$) must be closely spaced with some other state to
satisfy the solar (atmospheric) constraint. In category `B',
there are two pairs of almost degenerate neutrinos separated by
the larger gap (details later). Taking the LSND experiment into
consideration, `A' is incompatible with the data,
while `B' can be comfortably accommodated \cite{4nu}.

Two cases may arise if we work in `A'. The heaviest neutrino
could be (dominantly) sterile. Since it does not have any direct
$\Rsl$ couplings, its decay via mixing with other states will be
strongly suppressed, and we cannot set any bound on the $\Rsl$
couplings.  Therefore, this case is not interesting for our
purpose and we will not consider it in our analysis. On the
contrary, the heaviest state could be (dominantly) the
$\nu_\tau$, decaying via $\Rsl$ interactions.  Since
our analysis concerns, after all, the decay of a heavy
active neutrino into at least one lighter active state, this
latter case does not yield any different result from what we
obtain if we work in `B'. Hence, to simplify matters, we adopt
the framework `B' for our subsequent discussions.

In `B', which is a `two-pair' scenario, all four neutrinos participate
in the solution to the solar and atmospheric neutrino anomalies.  In
this case there are two allowed forms of the mass spectrum consisting
of two closely spaced pair of states (spacings $\Delta_{\rm atmos}$
and $\Delta_{\rm solar}$) with a larger separation ($\Delta$) between
the two pairs as shown in Fig.~1.  Inclusion of the LSND result fixes
the mass splitting $\Delta$ to $\Delta_{\rm LSND}$ \cite{4nu}.  It
needs to be mentioned that the oscillation results (excepting for the
MSW case) determine only the magnitude of the spacing between the
levels. Therefore, in addition to the two mass spectra shown in
Fig.~1, other possibilities obtained from these by exchanging the
members within any of the two closely spaced pairs or by interchanging
the relative ordering of the two pairs gives rise to spectra which are
equally acceptable.  As alluded to earlier, our results below are not
affected by such inversion of hierarchies.

\noindent
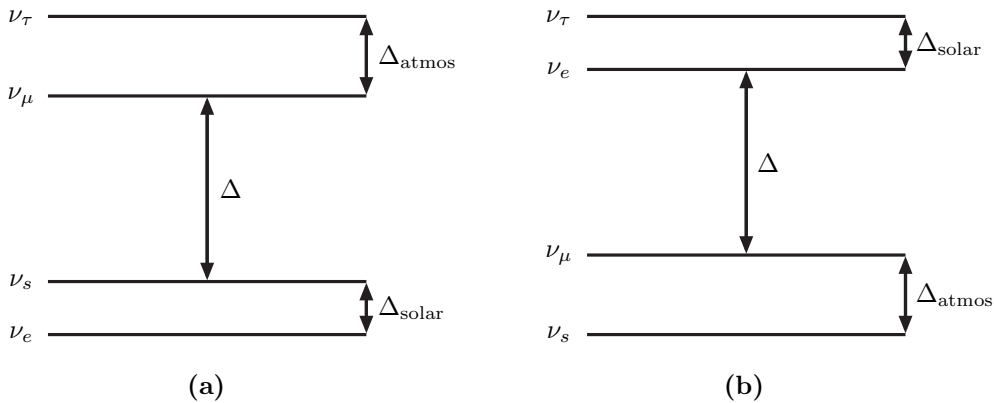
\begin{figure}[h]
\begin{center}
\begin{picture}(200,130)(0,0)
\SetWidth{1.2}
\Line(20,0)(140,0)
\Line(20,20)(140,20)
\Line(20,90)(140,90)
\Line(20,120)(140,120)
\LongArrow(80,22)(80,88)
\LongArrow(80,88)(80,22)
\LongArrow(140,2)(140,18)
\LongArrow(140,18)(140,2)
\LongArrow(140,92)(140,118)
\LongArrow(140,118)(140,92)
\Text(80,-20)[c]{\bf{(a)}}
\Text(10,0)[c]{$\nu_{e}$}
\Text(10,20)[c]{$\nu_{s}$}
\Text(10,90)[c]{$\nu_{\mu}$}
\Text(10,120)[c]{$\nu_{\tau}$}
\Text(145,10)[l]{$\Delta_{{\rm solar}}$}
\Text(145,105)[l]{$\Delta_{{\rm atmos}}$}
\Text(85,55)[l]{$\Delta$}
\end{picture}
%
\begin{picture}(200,100)(150,0)
\SetWidth{1.2}
\Line(170,0)(290,0)
\Line(170,30)(290,30)
\Line(170,100)(290,100)
\Line(170,120)(290,120)
\LongArrow(230,32)(230,98)
\LongArrow(230,98)(230,32)
\LongArrow(290,2)(290,28)
\LongArrow(290,28)(290,2)
\LongArrow(290,102)(290,118)
\LongArrow(290,118)(290,102)
\Text(230,-20)[c]{\bf{(b)}}
\Text(160,0)[c]{$\nu_{s}$}
\Text(160,30)[c]{$\nu_{\mu}$}
\Text(160,100)[c]{$\nu_{e}$}
\Text(160,120)[c]{$\nu_{\tau}$}
\Text(295,15)[l]{$\Delta_{{\rm atmos}}$}
\Text(295,110)[l]{$\Delta_{{\rm solar}}$}
\Text(235,65)[l]{$\Delta$}
\end{picture}
\end{center}
\vskip 20pt
\caption[]{\small\sf The two allowed forms of the neutrino mass
spectra from atmospheric and solar results. Note that the
spacings $\Delta_{{\rm atmos}}$ and $\Delta_{{\rm solar}}$ are
only indicative and not to scale. If the LSND results are also
included then $\Delta \equiv \Delta_{\rm LSND}$.}
\label{f:massdif} %
\end{figure}

In our subsequent work, pending confirmation from another independent
experiment, we do not impose the LSND constraint on the mass spectra.
Once we abandon the LSND results, both `A' and `B' are acceptable by
the remaining data. In any case, as noted before, except for omitting
the uninteresting case of the decaying heavy sterile neutrino in the
framework `A', we do not lose any generality by sticking to the option
`B'\footnote{We make a remark in passing that Big-Bang Nucleosynthesis
  (BBN) data constrain the mixing between an active and a sterile
  neutrino \cite{dolgov}. For example, the constraint on the
  additional number of neutrino species $\Delta N_\nu < 0.2$
  \cite{tytler} disfavours the large angle $\nu_\mu$-$\nu_s$ mixing
  solution to the atmospheric neutrino problem. A `safer' estimate
  $\Delta N_\nu < 1$ \cite{subir}, on the other hand, can accommodate
  an arbitrarily strong mixing between an active and a sterile state.
  For the present analysis we do not indulge ourselves further
  along these lines.}.  Thus, we assume that there are two
pairs of states with $\Delta m^2$ determined by the atmospheric and
solar neutrino results but we leave the splitting between these pairs
arbitrary, demanding only that consistency with the laboratory limits
be maintained. This permits us to consider mass splittings as large as
170 keV (but not more), which is the laboratory upper limit on
$m_{\nu_\mu}$, and check the cosmological implications if such heavy
neutrinos decay into lighter states via the $\Rsl$ interactions.  It
should be noted that, in this framework, the laboratory upper limit of
18.2 MeV on $m_{\nu_\tau}$ can never be reached, as Fig.~1a suggests
that $m_{\nu_\tau}$ can at most be a little above $m_{\nu_\mu}$.  The
generation indices for the decaying heavy neutrino and the lightest
neutrino produced as a decay product are chosen as the combinations
(3,1), (3,2), and (2,1) respectively.  We will neglect the mass of the
product neutrino.

\section {Radiative decay of a heavy neutrino} 

A heavy neutrino, which from now on we denote by $\nuH$, can decay
radiatively into a lighter neutrino (anti-neutrino), denoted by $\nu$
($\bar{\nu}$), through $\Delta L$ = 0 (2) penguin diagrams containing
$\Rsl$ couplings.  We discuss these two cases separately.

\subsection{The decay $\nuH \rightarrow \nu+ \gamma$}

The $\Delta L=0$ decay  of $\nuH$ to $\nu$ and a real 
photon occurs through penguin diagrams (see Fig.~2). 
\noindent
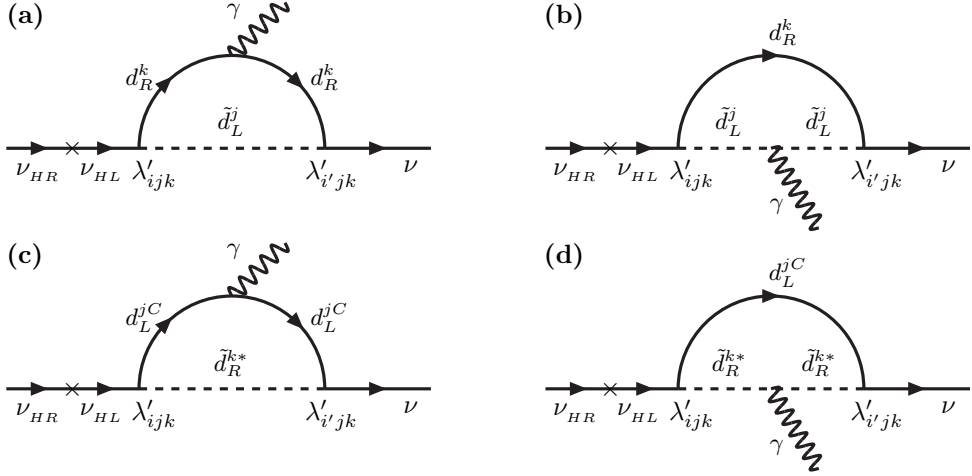
\begin{figure}[h]
\begin{center}
\begin{picture}(200,90)(0,0)
\SetWidth{1.2}
\ArrowLine(-10,0)(15,0)
\Text(15,0)[c]{$\times$}
\ArrowLine(15,0)(40,0)
\ArrowArcn(75,0)(35,180,90)
\ArrowArcn(75,0)(35,90,0)
\Photon(75,35)(95,55){3}{5}
\Text(76,52)[c]{\small{$\gamma$}}
\DashLine(110,0)(40,0){3}
\ArrowLine(110,0)(150,0)
\Text(-10,50)[l]{\bf{(a)}}
\Text(2,-9)[c]{$\nu_{_{HR}}$}
\Text(26,-9)[c]{$\nu_{_{HL}}$}
\Text(143,-7)[c]{$\nu$}
\Text(75,10)[c]{\small{$\tilde{d}_L^{j}$}}
\Text(35,27)[l]{\small{$d^k_{R}$}}
\Text(105,27)[l]{\small{$d^k_{R}$}}
\Text(46,-10)[c]{$\lambda'_{ijk}$}
\Text(113,-10)[c]{$\lambda'_{i'jk}$}
\end{picture}
%
\begin{picture}(200,90)(150,0)
\SetWidth{1.2}
\ArrowLine(140,0)(165,0)
\Text(165,0)[c]{$\times$}
\ArrowLine(165,0)(190,0)
\ArrowArcn(225,0)(35,180,0)
\Photon(225,0)(243,-30){3}{7}
\Text(228,-22)[c]{\small{$\gamma$}}
\DashLine(260,0)(190,0){3}
\ArrowLine(260,0)(300,0)
\Text(140,50)[l]{\bf{(b)}}
\Text(152,-9)[c]{$\nu_{_{HR}}$}
\Text(176,-9)[c]{$\nu_{_{HL}}$}
\Text(293,-7)[c]{$\nu$}
\Text(210,10)[c]{\small{$\tilde{d}_L^{j}$}}
\Text(244,10)[c]{\small{$\tilde{d}_L^{j}$}}
\Text(225,44)[l]{\small{$d^k_{R}$}}
\Text(196,-10)[c]{$\lambda'_{ijk}$}
\Text(266,-10)[c]{$\lambda'_{i'jk}$}
\end{picture}
\begin{picture}(200,90)(0,0)
\SetWidth{1.2}
\ArrowLine(-10,0)(15,0)
\Text(15,0)[c]{$\times$}
\ArrowLine(15,0)(40,0)
\ArrowArcn(75,0)(35,180,90)
\ArrowArcn(75,0)(35,90,0)
\Photon(75,35)(95,55){3}{5}
\Text(76,52)[c]{\small{$\gamma$}}
\DashLine(110,0)(40,0){3}
\ArrowLine(110,0)(150,0)
\Text(-10,50)[l]{\bf{(c)}}
\Text(2,-9)[c]{$\nu_{_{HR}}$}
\Text(26,-9)[c]{$\nu_{_{HL}}$}
\Text(143,-7)[c]{$\nu$}
\Text(75,10)[c]{\small{$\tilde{d}_R^{k*}$}}
\Text(35,27)[l]{\small{$d^{jC}_{L}$}}
\Text(105,27)[l]{\small{$d^{jC}_{L}$}}
\Text(46,-10)[c]{$\lambda'_{ijk}$}
\Text(113,-10)[c]{$\lambda'_{i'jk}$}
\end{picture}
%
\begin{picture}(200,90)(150,0)
\SetWidth{1.2}
\ArrowLine(140,0)(165,0)
\Text(165,0)[c]{$\times$}
\ArrowLine(165,0)(190,0)
\ArrowArcn(225,0)(35,180,0)
\Photon(225,0)(243,-30){3}{7}
\Text(228,-22)[c]{\small{$\gamma$}}
\DashLine(260,0)(190,0){3}
\ArrowLine(260,0)(300,0)
\Text(140,50)[l]{\bf{(d)}}
\Text(152,-9)[c]{$\nu_{_{HR}}$}
\Text(176,-9)[c]{$\nu_{_{HL}}$}
\Text(293,-7)[c]{$\nu$}
\Text(210,10)[c]{\small{$\tilde{d}_R^{k*}$}}
\Text(244,10)[c]{\small{$\tilde{d}_R^{k*}$}}
\Text(225,44)[l]{\small{$d^{jC}_{L}$}}
\Text(196,-10)[c]{$\lambda'_{ijk}$}
\Text(266,-10)[c]{$\lambda'_{i'jk}$}
\end{picture}
\end{center}
\vskip 20pt
\caption[]{\small\sf Penguin diagrams corresponding to the
$\Delta L = 0$ radiative decay of $\nuH$. }\label{f:raddel0} 
\end{figure}
Although the right-handed neutrinos do not couple in the $\Rsl$
superpotential, a mass insertion on the heavy neutrino line (which 
requires $\nuH$ to have Dirac-type mass) enables one
to write down the following effective Hamiltonian for this decay:
\beq
{\cal{H}}^{\Delta L=0}_{eff}=\frac{1}{2} 
{\cal{A}}_1\;\overline{\nu}\;P_R\;
\sigma^{\mu\nu}\;\nu_{_H}\;F_{\mu\nu}
\eeq
where
\beq
 {\cal{A}}_1 =  e\;\lambda'_{ijk}\lambda'_{i'jk} Q_d N_c 
{{m_{\nu_H}}\over{\ms^2}}
{{1}\over{16\pi^2}} 
\left({\cal{I}}_{1k} - {\cal{I}}_{1j} \right).
\label{exprl0}
\eeq
Here 
\beq 
{\cal{I}}_{1j}= {{1} \over{4(1-r_{d_j})^3}}\left(-1+r_{d_j}^2-2 r_{d_j}
\ln(r_{d_j}) \right) 
\eeq 
with $r_{d_j}=(m_{d_{j}}/\ms)^2$. In the above expression, and in all
subsequent discussions, we assume a common mass $\ms$ for whichever
scalar is exchanged.  $Q_d$ is the charge of the down-type quark
inside the loop. $N_c = 3$ is the colour factor.

\subsection{The decay $\nuH\rightarrow\bar\nu+\gamma$}

This $\Delta L=2$ decay of a $\nuH$ to $\bar\nu$ and a real photon
occurs through penguin diagrams (see Fig.~3).  Contrary to the
previous mode, for this decay to take place the decaying heavy
neutrino does not require to have a Dirac mass.
\noindent
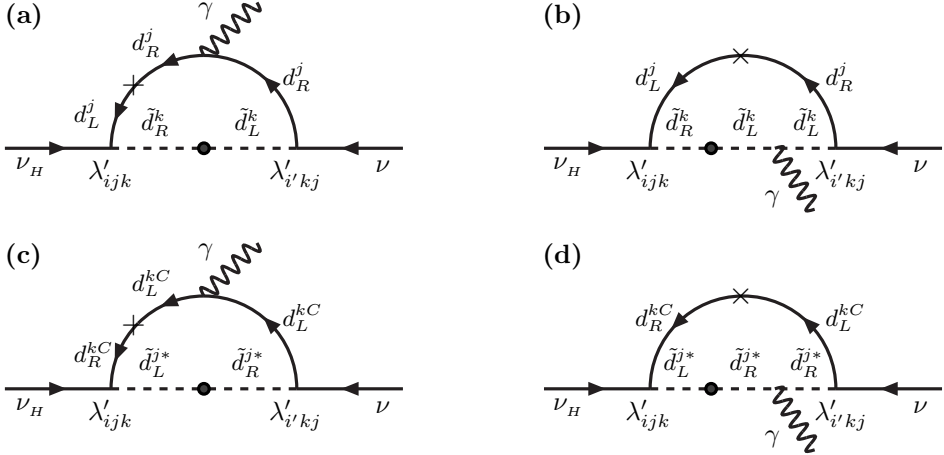
\begin{figure}[h]
\begin{center}
\begin{picture}(200,90)(0,0)
\SetWidth{1.2}
\ArrowLine(0,0)(40,0)
\ArrowArc(75,0)(35,0,90)
\ArrowArc(75,0)(35,135,180)
\ArrowArc(75,0)(35,90,135)
\Photon(75,35)(95,55){3}{5}
\Text(76,52)[c]{{$\gamma$}}
\DashLine(110,0)(40,0){3}
\ArrowLine(150,0)(110,0)
\GCirc(75,0){2}{.25}
\Text(0,50)[l]{\bf{(a)}}
\Text(10,-7)[c]{$\nu_{_H}$}
\Text(143,-7)[c]{$\nu$}
\Text(57,10)[c]{\small{$\tilde{d}_R^{k}$}}
\Text(92,10)[c]{\small{$\tilde{d}_L^{k}$}}
\Text(44,24)[l]{\large{$+$}}
\Text(26,13)[l]{\small{$d^j_{L}$}}
\Text(48,40)[l]{\small{$d^j_{R}$}}
\Text(105,27)[l]{\small{$d^j_{R}$}}
\Text(40,-10)[c]{$\lambda'_{ijk}$}
\Text(110,-10)[c]{$\lambda'_{i'kj}$}
\end{picture}
%
\begin{picture}(200,90)(150,0)
\SetWidth{1.2}
\ArrowLine(150,0)(190,0)
\ArrowArc(225,0)(35,0,90)
\ArrowArc(225,0)(35,90,180)
\Photon(237,0)(252,-23){3}{5}
\Text(237,-19)[c]{{$\gamma$}}
\DashLine(260,0)(190,0){3}
\ArrowLine(300,0)(260,0)
\Text(225,35)[c]{\large{$\times$}}
\GCirc(213,0){2}{.25}
\Text(150,50)[l]{\bf{(b)}}
\Text(160,-7)[c]{$\nu_{_H}$}
\Text(293,-7)[c]{$\nu$}
\Text(202,10)[c]{\small{$\tilde{d}_R^{k}$}}
\Text(227,10)[c]{\small{$\tilde{d}_L^{k}$}}
\Text(250,10)[c]{\small{$\tilde{d}_L^{k}$}}
\Text(185,27)[l]{\small{$d^j_{L}$}}
\Text(257,27)[l]{\small{$d^j_{R}$}}
\Text(190,-10)[c]{$\lambda'_{ijk}$}
\Text(263,-10)[c]{$\lambda'_{i'kj}$}
\end{picture}
\begin{picture}(200,90)(0,0)
\SetWidth{1.2}
\ArrowLine(0,0)(40,0)
\ArrowArc(75,0)(35,0,90)
\ArrowArc(75,0)(35,135,180)
\ArrowArc(75,0)(35,90,135)
\Photon(75,35)(95,55){3}{5}
\Text(76,52)[c]{{$\gamma$}}
\DashLine(110,0)(40,0){3}
\ArrowLine(150,0)(110,0)
\GCirc(75,0){2}{.25}
\Text(0,50)[l]{\bf{(c)}}
\Text(10,-7)[c]{$\nu_{_H}$}
\Text(143,-7)[c]{$\nu$}
\Text(57,10)[c]{\small{$\tilde{d}_L^{j*}$}}
\Text(92,10)[c]{\small{$\tilde{d}_R^{j*}$}}
\Text(44,24)[l]{\large{$+$}}
\Text(26,13)[l]{\small{$d^{kC}_{R}$}}
\Text(48,40)[l]{\small{$d^{kC}_{L}$}}
\Text(105,27)[l]{\small{$d^{kC}_{L}$}}
\Text(40,-10)[c]{$\lambda'_{ijk}$}
\Text(110,-10)[c]{$\lambda'_{i'kj}$}
\end{picture}
%
\begin{picture}(200,90)(150,0)
\SetWidth{1.2}
\ArrowLine(150,0)(190,0)
\ArrowArc(225,0)(35,0,90)
\ArrowArc(225,0)(35,90,180)
\Photon(237,0)(252,-23){3}{5}
\Text(237,-19)[c]{{$\gamma$}}
\DashLine(260,0)(190,0){3}
\ArrowLine(300,0)(260,0)
\Text(225,35)[c]{\large{$\times$}}
\GCirc(213,0){2}{.25}
\Text(150,50)[l]{\bf{(d)}}
\Text(160,-7)[c]{$\nu_{_H}$}
\Text(293,-7)[c]{$\nu$}
\Text(202,10)[c]{\small{$\tilde{d}_L^{j*}$}}
\Text(227,10)[c]{\small{$\tilde{d}_R^{j*}$}}
\Text(250,10)[c]{\small{$\tilde{d}_R^{j*}$}}
\Text(185,27)[l]{\small{$d^{kC}_{R}$}}
\Text(257,27)[l]{\small{$d^{kC}_{L}$}}
\Text(190,-10)[c]{$\lambda'_{ijk}$}
\Text(263,-10)[c]{$\lambda'_{i'kj}$}
\end{picture}
\end{center}
\vskip 20pt
\caption[]{\small\sf Penguin diagrams corresponding to the
$\Delta L = 2$ radiative decay of $\nuH$.}\label{f:raddel2} 
\end{figure}
The effective Hamiltonian for this decay is given by
\beq
{\cal{H}}^{\Delta L=2}_{eff}=\frac{1}{2}{\cal{A}}_2\;
\overline{\nu^c}\;P_L\;\sigma^{\mu\nu}\;\nuH\;F_{\mu\nu}
\eeq
where
\beq
{\cal{A}}_2=  e\;\lambda'_{ijk} \lambda'_{i'kj} Q_d N_c
{{m_{d_j} m_{d_k} } \over {\ms^3} }
{{1}\over{8\pi^2}}  
\left( {\cal{I}}_{2k}-{\cal{I}}_{2j} \right).
\label{exprl2}
\eeq
Here
\beq
{\cal{I}}_{2j}= {{1} \over{(1-r_{d_j})^2}}
\left(1 -  r_{d_j}  + \ln(r_{d_j}) \right).
\eeq

\vskip 10pt
\noindent 

As is seen from Fig. 3 this mode involves the mixing between left-
and right-type $d$-squarks. This has been approximated by $\Delta m^2
({\rm LR}) = (A - \mu \tan \beta)~ m_d \sim \ms m_d$ in the above
expression.

It follows from Eqs. (\ref{exprl0}) and (\ref{exprl2}) that for the
special case when $j = k$, the radiative decay amplitude vanishes
identically. A look at Figs. 2 and 3 reveals that for each diagram in
which a photon is attached to a particle (fermion or scalar) of
generation $j$ there exists a similar diagram in which the photon
couples to the corresponding anti-particle of generation $k$ -- hence
the cancellation when the generations match.

In the above we have considered processes which are triggered by
non-vanishing $\lambda'$ couplings. These processes could also be
driven by $\lambda$-type couplings.  The transition can be readily
performed in Figs. 2 and 3 by replacing the internal down-type
quarks with charged leptons and the down-type squarks with
sleptons of the corresponding generations.  Indeed, $N_c$ would
be unity in this situation.

\section{The invisible decay of $\nuH$ into three light neutrinos}

The other kinematically allowed decay mode of $\nuH$ is into three
neutrinos/anti-neutrinos. Such a decay proceeds via a set of
$Z$-mediated penguin diagrams (see Fig.~4). The choice of $\lambda'$
(or $\lambda$) couplings determines the flavour of one final state
neutrino. In fact, for the aforenoted case of $j =k$ leading to
vanishing radiative decays, such three-body invisible channels
constitute the only decay modes of $\nuH$. Since the $Z$ boson, unlike
the photon, couples to fermions (sfermions) and their conjugates with
different strengths, the cancellation observed for a radiative decay
process does not occur for a $Z$-mediated penguin. As we will see
later, the decay widths in these three-body modes are much smaller
than the two-body radiative decay widths.  In the following, we shall
consider only the $\Delta L=0$ penguins. The $\Delta L=2$ diagrams
suffer a suppression from left-right sfermion mixing, and we will not
consider them in our subsequent discussions as their numerical impact
is insignificant. It is also possible to drive this decay through box
graphs involving the products of $\lambda$ couplings, but their
contribution to the same order as that of the penguins, we have
checked, is insignificant.
\begin{figure}[htbp]
\begin{center}
\begin{picture}(200,90)(0,0)
\SetWidth{1.2}
\ArrowLine(0,0)(40,0)
\ArrowArcn(75,0)(35,180,90)
\ArrowArcn(75,0)(35,90,0)
\Photon(75,35)(95,55){3}{5}
\Text(76,52)[c]{\small{$Z$}}
\ArrowLine(115,75)(95,55)
\Text(119,68)[c]{$\nu$}
\ArrowLine(95,55)(115,35)
\DashLine(110,0)(40,0){3}
\ArrowLine(110,0)(150,0)
\Text(-12,50)[l]{\bf{(a)}}
\Text(10,-7)[c]{$\nu_{_H}$}
\Text(119,44)[c]{$\nu$}
\Text(10,-7)[c]{$\nu_{_H}$}
\Text(143,-7)[c]{$\nu$}
\Text(75,10)[c]{\small{$\tilde{d}_L^{j}$}}
\Text(35,27)[l]{\small{$d^k_{R}$}}
\Text(105,27)[l]{\small{$d^k_{R}$}}
\Text(40,-10)[c]{$\lambda'_{ijk}$}
\Text(110,-10)[c]{$\lambda'_{i'jk}$}
\end{picture}
%
\begin{picture}(200,90)(150,0)
\SetWidth{1.2}
\ArrowLine(150,0)(190,0)
\ArrowArcn(225,0)(35,180,0)
\Photon(225,0)(252,-35){3}{7}
\Text(230,-22)[c]{\small{$Z$}}
\ArrowLine(272,-15)(252,-35)
\ArrowLine(252,-35)(272,-55)
\DashLine(260,0)(190,0){3}
\ArrowLine(260,0)(300,0)
\Text(138,50)[l]{\bf{(b)}}
\Text(278,-22)[c]{$\nu$}
\Text(160,-7)[c]{$\nu_{_H}$}
\Text(278,-48)[c]{$\nu$}
\Text(293,-7)[c]{$\nu$}
\Text(210,10)[c]{\small{$\tilde{d}_L^{j}$}}
\Text(244,10)[c]{\small{$\tilde{d}_L^{j}$}}
\Text(225,44)[l]{\small{$d^k_{R}$}}
\Text(190,-10)[c]{$\lambda'_{ijk}$}
\Text(260,-10)[c]{$\lambda'_{i'jk}$}
\end{picture}
\begin{picture}(200,130)(0,0)
\SetWidth{1.2}
\ArrowLine(0,0)(30,0)
\ArrowLine(30,0)(65,0)
\ArrowArcn(100,0)(35,180,0)
\Photon(30,0)(45,50){3}{10}
\Text(69,63)[c]{$\nu$}
\ArrowLine(65,70)(45,50)
\Text(27,25)[c]{\small{$Z$}}
\ArrowLine(45,50)(65,30)
\DashLine(135,0)(65,0){3}
\ArrowLine(135,0)(175,0)
\Text(0,50)[l]{\bf{(c)}}
\Text(10,-7)[c]{$\nu_{_H}$}
\Text(168,-7)[c]{$\nu$}
\Text(100,10)[c]{\small{$\tilde{d}_L^{j}$}}
\Text(100,43)[l]{\small{$d^k_{R}$}}
\Text(47,-9)[c]{$\nu_{_H}$}
\Text(69,39)[c]{$\nu$}
\Text(68,-10)[c]{$\lambda'_{ijk}$}
\Text(135,-10)[c]{$\lambda'_{i'jk}$}
\end{picture}
%
\begin{picture}(220,130)(150,0)
\SetWidth{1.2}
\ArrowLine(150,0)(190,0)
\ArrowArcn(225,0)(35,180,0)
\Photon(290,0)(305,50){3}{10}
\ArrowLine(325,70)(305,50)
\ArrowLine(305,50)(325,30)
\DashLine(260,0)(190,0){3}
\Text(225,10)[c]{\small{$\tilde{d}_L^{j}$}}
\ArrowLine(260,0)(290,0)
\ArrowLine(290,0)(325,0)
\Text(276,-7)[c]{$\nu$}
\Text(149,50)[l]{\bf{(d)}}
\Text(160,-7)[c]{$\nu_{_H}$}
\Text(289,25)[c]{\small{$Z$}}
\Text(329,63)[c]{$\nu$}
\Text(330,38)[c]{$\nu$}
\Text(160,-7)[c]{$\nu_{_H}$}
\Text(320,-7)[c]{$\nu$}
\Text(225,44)[l]{\small{$d^k_{R}$}}
\Text(190,-10)[c]{$\lambda'_{ijk}$}
\Text(260,-10)[c]{$\lambda'_{i'jk}$}
\end{picture}
\vskip 40pt
\begin{picture}(200,90)(0,0)
\SetWidth{1.2}
\ArrowLine(0,0)(40,0)
\ArrowArcn(75,0)(35,180,90)
\ArrowArcn(75,0)(35,90,0)
\Photon(75,35)(95,55){3}{5}
\Text(76,52)[c]{\small{$Z$}}
\ArrowLine(115,75)(95,55)
\Text(119,68)[c]{$\nu$}
\ArrowLine(95,55)(115,35)
\DashLine(110,0)(40,0){3}
\ArrowLine(110,0)(150,0)
\Text(-8,50)[l]{\bf{(e)}}
\Text(10,-7)[c]{$\nu_{_H}$}
\Text(119,44)[c]{$\nu$}
\Text(10,-7)[c]{$\nu_{_H}$}
\Text(143,-7)[c]{$\nu$}
\Text(75,10)[c]{\small{$\tilde{d}_R^{k*}$}}
\Text(35,27)[l]{\small{$d^{jC}_{L}$}}
\Text(105,27)[l]{\small{$d^{jC}_{L}$}}
\Text(40,-10)[c]{$\lambda'_{ijk}$}
\Text(110,-10)[c]{$\lambda'_{i'jk}$}
\end{picture}
%
\begin{picture}(200,90)(150,0)
\SetWidth{1.2}
\ArrowLine(150,0)(190,0)
\ArrowArcn(225,0)(35,180,0)
\Photon(225,0)(252,-35){3}{7}
\Text(230,-22)[c]{\small{$Z$}}
\ArrowLine(272,-15)(252,-35)
\ArrowLine(252,-35)(272,-55)
\DashLine(260,0)(190,0){3}
\ArrowLine(260,0)(300,0)
\Text(140,50)[l]{\bf{(f)}}
\Text(278,-22)[c]{$\nu$}
\Text(160,-7)[c]{$\nu_{_H}$}
\Text(278,-48)[c]{$\nu$}
\Text(293,-7)[c]{$\nu$}
\Text(210,10)[c]{\small{$\tilde{d}_R^{k*}$}}
\Text(244,10)[c]{\small{$\tilde{d}_R^{k*}$}}
\Text(225,44)[l]{\small{$d^{jC}_{L}$}}
\Text(190,-10)[c]{$\lambda'_{ijk}$}
\Text(260,-10)[c]{$\lambda'_{i'jk}$}
\end{picture}
\begin{picture}(200,130)(0,0)
\SetWidth{1.2}
\ArrowLine(0,0)(30,0)
\ArrowLine(30,0)(65,0)
\ArrowArcn(100,0)(35,180,0)
\Photon(30,0)(45,50){3}{10}
\Text(69,63)[c]{$\nu$}
\ArrowLine(65,70)(45,50)
\Text(27,25)[c]{\small{$Z$}}
\ArrowLine(45,50)(65,30)
\DashLine(135,0)(65,0){3}
\ArrowLine(135,0)(175,0)
\Text(2,50)[l]{\bf{(g)}}
\Text(10,-7)[c]{$\nu_{_H}$}
\Text(168,-7)[c]{$\nu$}
\Text(100,10)[c]{\small{$\tilde{d}_R^{k*}$}}
\Text(100,43)[l]{\small{$d^{jC}_{L}$}}
\Text(47,-9)[c]{$\nu_{_H}$}
\Text(69,39)[c]{$\nu$}
\Text(68,-10)[c]{$\lambda'_{ijk}$}
\Text(135,-10)[c]{$\lambda'_{i'jk}$}
\end{picture}
%
\begin{picture}(220,130)(150,0)
\SetWidth{1.2}
\ArrowLine(150,0)(190,0)
\ArrowArcn(225,0)(35,180,0)
\Photon(290,0)(305,50){3}{10}
\ArrowLine(325,70)(305,50)
\ArrowLine(305,50)(325,30)
\DashLine(260,0)(190,0){3}
\Text(225,10)[c]{\small{$\tilde{d}_R^{k*}$}}
\ArrowLine(260,0)(290,0)
\ArrowLine(290,0)(325,0)
\Text(276,-7)[c]{$\nu$}
\Text(150,50)[l]{\bf{(h)}}
\Text(160,-7)[c]{$\nu_{_H}$}
\Text(289,25)[c]{\small{$Z$}}
\Text(329,63)[c]{$\nu$}
\Text(330,38)[c]{$\nu$}
\Text(160,-7)[c]{$\nu_{_H}$}
\Text(320,-7)[c]{$\nu$}
\Text(225,44)[l]{\small{$d^{jC}_{L}$}}
\Text(190,-10)[c]{$\lambda'_{ijk}$}
\Text(260,-10)[c]{$\lambda'_{i'jk}$}
\end{picture}
\end{center}
\caption[]{\small\sf The $Z$-mediated penguin graphs for the ($\Delta
L =0$) $\nuH \rightarrow \nu \nu \bar{\nu}$ decay.}\label{f:Zdel0} %
\end{figure}
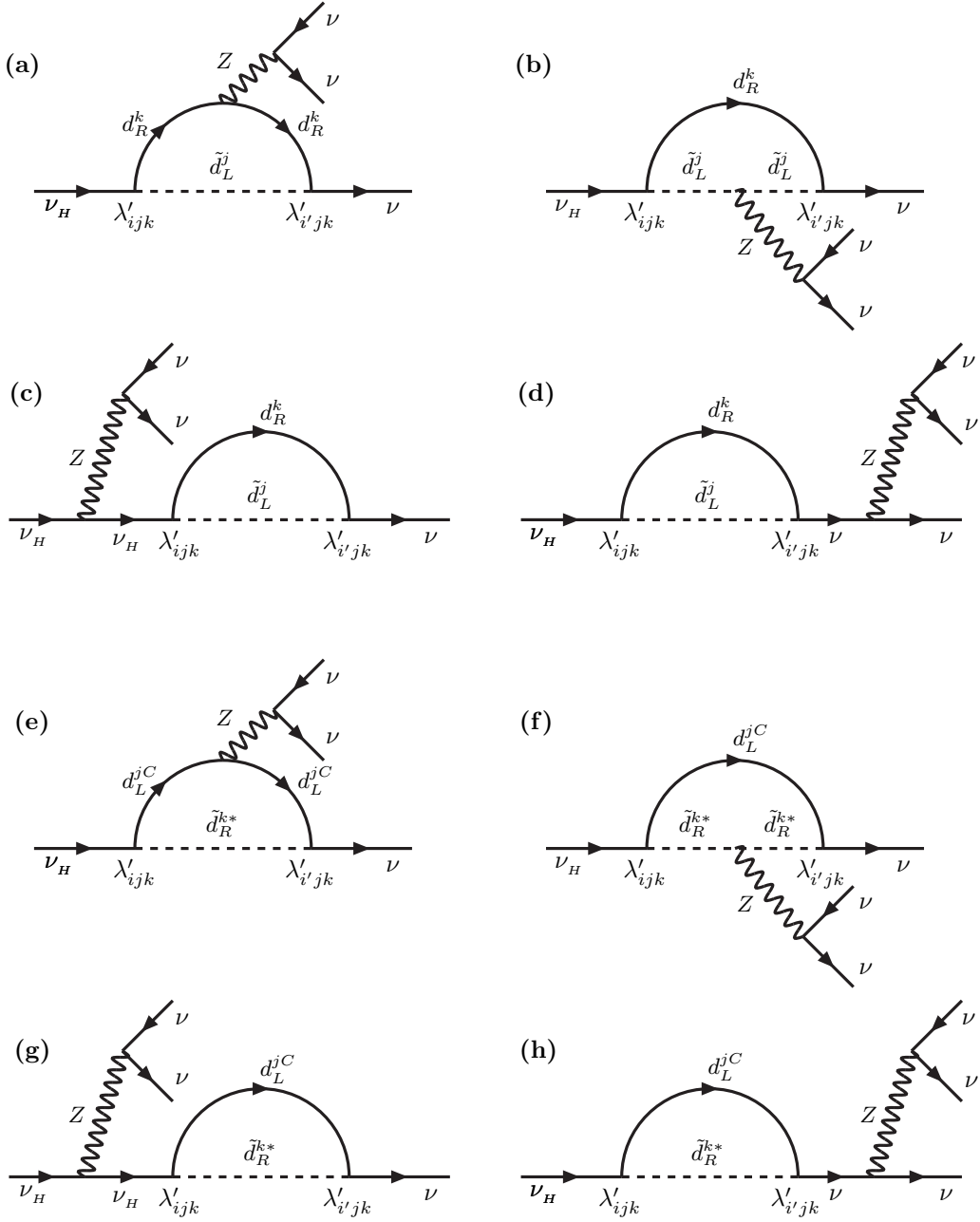

For simplicity, we present the expressions for the $\Delta L = 0$
penguin for those cases in which the internal fermion and the sfermion
are of the same type ({i.e.}, $j=k$) and the heavy neutrino mass is
negligible in comparison with the internal fermion masses. In such a
situation the one-loop effective $Z\nu\nuH$ vertex can be written as
$-i\frac{g}{\cos\theta_W} \delta a_L \gamma_{\mu} P_L$, 
where,
\beq
\delta a_L= N_c \frac{\lambda'_{ijj}\lambda'_{i'jj}}{16 \pi^2} \left[
\frac{r_{d_j}}{1-r_{d_j}} + \frac{r_{d_j}
\ln(r_{d_j})}{(1-r_{d_j})^2}\right].  
\eeq

\section{Cosmological constraints}\label{cosmos}

Cosmology sets tight constraints on allowed neutrino masses and
associated lifetimes. For a neutrino of a certain mass, any decay mode
driven by $\Rsl$ interactions has a lifetime determined by the
couplings and the exchanged particle masses. It needs to be checked
whether such decays are consistent with the cosmological requirements
(discussed below) or whether one obtains new bounds on the $\Rsl$
couplings. We discuss the different modes in the following
subsections. In order to permit easy comparison with the existing
bounds on such couplings, we choose, as in the literature, all
exchanged supersymmetric scalar particle (squark or slepton) masses to
be 100 GeV in the discussion below. The bounds corresponding to any
other sfermion mass can be readily obtained by using the formulae
given in this paper (although just by scaling arguments one can derive
these numbers to a very good approximation).

Though we are examining {\em neutrino decays}, the case of a
cosmologically stable neutrino plays an important role. If the
lifetime of the neutrino is larger than the age of the universe
($t_0$) \cite{t0}, then it is cosmologically stable.  They must either
be rather light (the Cowsik-McClelland bound \cite{cowsik}) or
comparatively heavy (the Lee-Weinberg bound \cite{lee}). The allowed
values for a stable neutrino mass are~\cite{numass}
\beq m_{\nu} \leq 92h^2 \Omega_0 ~{\rm
eV},\;\;\;{\rm or}\;\;\; m_{\nu} \geq \frac{2}{\sqrt{h^2 \Omega_0}}
~{\rm GeV}
\label{stbound}
\eeq 
Here, $\Omega_0$ is the ratio of the present energy density to the
critical energy density and $h$ is the normalized Hubble expansion
rate. $h$ is defined as $H_0=100 h ~{\rm km}~{\rm s}^{-1}~{\rm
Mpc}^{-1}$, where $H_0$ is the Hubble expansion rate. Present data
indicate that $h^2 \Omega_0$ is smaller than unity.  Since we are
considering neutrinos of mass 100 keV or less, the second inequality
in Eq.~(\ref{stbound}) is not of relevance for us.  For the sake of
discussion we take the first limit to be 45 eV; its exact value
depends on the precise magnitude of the combination $h^2 \Omega_0$.
Hence, a neutrino of mass 45 eV or more must necessarily be
cosmologically unstable.

\subsection{The $\nuH \rightarrow \nu (\bar{\nu}) + \gamma$ mode}

The final state photons in radiative decays of neutrinos are
subject to constraints from the cosmic microwave background
radiation (CMBR) spectrum, primordial nucleosynthesis, etc.  For
the $\Delta L=2$ radiative process $\nuH \rightarrow
\bar{\nu} + \gamma$, the neutrino decay lifetime is
\beq
\tau_{\nu}({\rm s}) \sim \frac{1}{[m_{\nu_{H}}({\rm eV})]^3} 
\frac{1}{[\lambda'\lambda']^{2}}\; f_1,
\label{rl2}
\eeq
where $f_1$ lies in the range  $4.8\times10^{19}$ --- $1.4\times10^{25}$
depending on the masses of the quarks and squarks exchanged in
the loops. [$\lambda'\lambda'$] denotes the product of the relevant
$\Rsl$ couplings. When the $\lambda$ couplings are involved,
$f_1$ lies between $1.0\times10^{21}$ --- $3.6\times10^{27}$.

For the sake of numerics, we choose a reference value of $m_{\nu_{H}}
=$ 45 eV. Since perturbativity requires all $\Rsl$ couplings be
smaller than unity, the lifetime of the neutrino from Eq.~({\ref{rl2})
is larger than $5.3\times10^{14}$s, comparable to $t_{rec} \sim 3
\times 10^{12}$s (the recombination epoch). Now $\tau_\nu \sim
t_{rec}$ for a 45 eV neutrino is not admissible from cosmological
requirements on the following grounds. If $t_0 > \tau_\nu > t_{rec}$,
then the photons produced in the radiative decay undergo redshift and
contribute to the diffuse photon background since after recombination
no charged particles are available to scatter them. This sets a bound
\cite{kt}:
\beq \left(\frac{m_\nu}{1 ~{\rm
eV}}\right)^{3/2} \leq
3\times10^{9}~(\Omega_0 h^2)^{-1/2} ~\left(\frac{\tau_\nu}{1
~{\rm s}}\right)^{-1}. 
\label{eq:trec}
\eeq 
A 45 eV neutrino cannot satisfy this constraint if its lifetime is as
estimated above.

The remaining possibility then is $\tau_{\nu} > t_0$.  These photons
would simply get superimposed on the diffuse photon background. The
experimental data, in this situation, yield the bound: 
\beq
\left(\frac{m_\nu}{1 ~{\rm eV}}\right)\leq 10^{-23}~
\left(\frac{\tau_\nu}{1~{\rm s}}\right). 
\label{eq:t0}
\eeq 
The above equation can be translated to upper bounds on the $\Rsl$
couplings. In Table~\ref{table1}, we have collected these new upper
bounds emerging from the $\Delta L=2$ radiative decay for a 45 eV
decaying neutrino. Side by side, we have displayed the existing
constraints on these couplings. We have presented only those cases for
which our new bounds are more stringent than the existing ones
and have rounded off these numbers to the first significant digit.
\begin{table}[h]
\begin{center}
 $$
\begin{array}{ccc} 
\hline\hline
{\rm ~~Combinations~~} & {\rm ~~Existing~~bounds~~} &
 {\rm ~~New ~~bounds~~}\\
\hline\hline
\lambda'_{313} \lambda'_{131} & 1 \times 10^{-3} & 1 \times10^{-4}\\
\lambda'_{321} \lambda'_{112} & 7 \times 10^{-3} & 6 \times10^{-3}\\
\lambda'_{323} \lambda'_{132} & 1 \times 10^{-1} & 1 \times10^{-5}\\
\lambda'_{331} \lambda'_{113} & 3 \times 10^{-3} & 1 \times10^{-4}\\
\lambda'_{332} \lambda'_{123} & 7 \times 10^{-3} & 1 \times10^{-5}\\

\lambda'_{313} \lambda'_{231} & 9 \times 10^{-3} & 1 \times10^{-4}\\
\lambda'_{321} \lambda'_{212} & 2 \times 10^{-2} & 6 \times10^{-3}\\
\lambda'_{323} \lambda'_{232} & 1 \times 10^{-1} & 1 \times10^{-5}\\
\lambda'_{331} \lambda'_{213} & 9 \times 10^{-3} & 1 \times10^{-4}\\
\lambda'_{332} \lambda'_{223} & 1 \times 10^{-2} & 1 \times10^{-5}\\

\lambda'_{213} \lambda'_{131} & 1 \times 10^{-3} & 1 \times10^{-4}\\
\lambda'_{223} \lambda'_{132} & 2 \times 10^{-2} & 1 \times10^{-5}\\
\lambda'_{231} \lambda'_{113} & 4 \times 10^{-3} & 1 \times10^{-4}\\
\lambda'_{232} \lambda'_{123} & 1 \times 10^{-2} & 1 \times10^{-5}\\
\hline
\lambda_{323} \lambda_{132} &   3 \times 10^{-3} & 5 \times10^{-5}\\ 
\lambda_{323} \lambda_{232} &   3 \times 10^{-3} & 5 \times10^{-5}\\ 
\lambda_{232} \lambda_{123} &   2 \times 10^{-3} & 5 \times10^{-5}\\ 
\hline\hline
\end{array}
$$
\end{center}
\caption[]{\small\sf New upper bounds on products of different
$\lambda'$ and $\lambda$ couplings for a heavy neutrino mass of 45 eV
and for $\Delta L = 2$ radiative decay. The existing bounds are
obtained by multiplying the upper bounds on the individual couplings
\cite{review}.}\label{table1}
\end{table}

Alternatively, one can turn the argument around and find the
maximum mass of the decaying heavy neutrino that is acceptable if
the $\Rsl$ couplings achieve their present experimental upper
limits. We have presented these bounds in Table~\ref{table2}
using the most dominant mode --- the $\Delta L = 2$ radiative
decay.

Now consider a neutrino of mass of 100 keV (this is near the maximum
limit of 170 keV, as we argued in section 2). Since $\Rsl$-couplings
must be smaller than unity, one obtains, from Eq.~(\ref{rl2}),
$\tau_\nu \gtap\: 4.8 \times 10^4$ s. But dumping of extra photons
by such massive neutrinos decaying at a time close to the
thermalization epoch and much after nucleosynthesis is not
cosmologically acceptable
\cite{GeRou,othercosm}, the relevant bound from the black-body
nature of the CMBR being:
\beq
\left(\frac{m_\nu}{1 ~{\rm eV}}\right)\leq 10^{7}~
\left(\frac{\tau_\nu}{1~{\rm s}}\right)^{-1/2}. 
\eeq

\begin{table}[h]
\begin{center}
 $$
\begin{array}{ccc} 
\hline\hline
{\rm ~~Combinations~~} & {\rm ~~Existing~~bounds~~} &
{~~{\rm Corresponding}~~{\rm maximum}}\\
& &
{~~m_{\nu_H}~~{\rm in ~~eV}}\\
\hline\hline

\lambda'_{313} \lambda'_{131} & 1 \times 10^{-3} & 15 \\
\lambda'_{321} \lambda'_{112} & 7 \times 10^{-3} & 40 \\
\lambda'_{323} \lambda'_{132} & 1 \times 10^{-1} & 0.4 \\
\lambda'_{331} \lambda'_{113} & 3 \times 10^{-3} & 8 \\
\lambda'_{332} \lambda'_{123} & 7 \times 10^{-3} & 2 \\

\lambda'_{313} \lambda'_{231} & 9 \times 10^{-3} & 5 \\
\lambda'_{321} \lambda'_{212} & 2 \times 10^{-2} & 26 \\
\lambda'_{323} \lambda'_{232} & 1 \times 10^{-1} & 0.4 \\
\lambda'_{331} \lambda'_{213} & 9 \times 10^{-3} & 5 \\
\lambda'_{332} \lambda'_{223} & 1 \times 10^{-2} & 1 \\

\lambda'_{213} \lambda'_{131} & 1 \times 10^{-3} & 15 \\
\lambda'_{223} \lambda'_{132} & 2 \times 10^{-2} & 1 \\
\lambda'_{231} \lambda'_{113} & 4 \times 10^{-3} & 8 \\
\lambda'_{232} \lambda'_{123} & 1 \times 10^{-2} & 1 \\
\hline
\lambda_{323} \lambda_{132} &   3 \times 10^{-3} & 6 \\
\lambda_{323} \lambda_{232} &   3 \times 10^{-3} & 6 \\
\lambda_{232} \lambda_{123} &   2 \times 10^{-3} & 7 \\ 
\hline\hline
\end{array}
$$
\end{center}
\caption[]{\small\sf Maximum allowed values of the heavy neutrino
mass decaying in the $\Delta L = 2$ radiative mode, based on the
assumption that the $\lambda'$ and $\lambda$ couplings achieve
their existing upper limits. These existing limits are as in
column 2 of Table 1. Recall that in Table 1 we have set new upper
limits on the relevant product couplings assuming a mass of 45
eV for the decaying neutrino.}
\label{table2}
\end{table}

For the $\Delta L=0$ radiative decay $\nuH \rightarrow \nu +
\gamma$, the amplitude is proportional to the Dirac mass of the
decaying neutrino. So the behaviour for the lifetime is as
follows: 
\beq 
\tau_{\nu}({\rm s}) \sim \frac{1}{[m_{\nu_{H}}({\rm eV})]^5}
\frac{1}{[\lambda'\lambda']^{2}}\; f_2,
\label{rl0}
\eeq
where $f_2$ lies between $2.1\times10^{40}$ --- $1.7\times10^{45}$.  
For the $\lambda$ couplings, $f_2$ lies in
the range $4.2\times10^{41}$ --- $9.9\times10^{45}$. For a 45 eV
decaying neutrino, Eq.~(\ref{rl0}) implies a lifetime larger than
the age of the universe even after setting the relevant $\Rsl$
couplings at their experimental upper limits. Thus, $R$-parity
violating couplings cannot be constrained from this process and
if the only non-vanishing $\Rsl$ couplings turn out
to be those that drive this $\Delta L = 0$ decay, then a neutrino
of mass 45 eV, in order to be cosmologically acceptable, must
have other interactions resulting in a faster decay.  It is easy
to convince oneself from the scaling argument that for a decaying
neutrino mass as high as 100 keV, the lifetime is either larger
than the age of the universe or, if less, certainly larger than
$t_{rec}$, the recombination epoch.  In the latter case, the
neutrino mass and radiative decay lifetime must satisfy relation
(\ref{eq:trec}), which is clearly not possible. Therefore, a 100
keV neutrino decaying via $\Rsl$ couplings {\em alone} is
inadmissible.

\subsection{The $\nuH \rightarrow \nu + \nu + \bar{\nu}$ mode} 

Usually the radiative decay mode dominates over the other loop decay
modes and consequently the latter are of any numerical relevance only
if the former is absent. The radiative decay amplitude involves the
combinations $\lambda'_{ijk} \lambda'_{i'kj}$ for the $\nuH
\rightarrow \bar{\nu} + \gamma$ mode (see Eq.~(\ref{exprl2})) and
$\lambda'_{ijk} \lambda'_{i'jk}$ for the $\nuH \rightarrow \nu +
\gamma$ process (see Eq.~(\ref{exprl0})). In the special case $j = k$,
the amplitude vanishes identically in each case, as can be easily seen
from Eqs.~(\ref{exprl0}) and (\ref{exprl2}). The same situation
prevails for the $\lambda $ couplings. Therefore, for bounds on the
$\lambda'_{ijj}\lambda'_{i'jj}$ (or $\lambda_{ijj}\lambda_{i'jj}$)
combinations we must turn to other processes. It is in these
situations that the present mode becomes of relevance.

For the $\Delta L=0$ $Z$-mediated $\nu_{_H}\to\nu\nu\bar{\nu}$ process
(the two neutrinos coupled to the $Z$-vertex must have the same
flavour) the lifetime is given by
\beq \tau_{\nu}({\rm s}) \sim
\frac{1}{[m_{\nu_{H}}({\rm eV})]^5} \frac{1}{[\lambda'\lambda']^{2}}\;
f_{3}.
\label{nl0}
\eeq 
Here $f_{3}$ lies in the range $10^{42}$---$10^{52}$. For the
$\lambda$ couplings, the range is $10^{44}$---$10^{57}$.  It is
seen from Eq.~(\ref{nl0}) that for a 45 eV (or even 100 keV)
neutrino, the lifetime for this process is always larger than the
age of the universe, $t_0$. Therefore, the $\Rsl$-couplings
cannot be constrained from this process for such neutrinos.

\section{Conclusions and Discussions}

We conclude by highlighting the salient features of our analysis.
The main thrust is to constrain from cosmological considerations
those $\Rsl$ couplings which trigger a heavy neutrino decay
either radiatively into a lighter neutrino or into three lighter
neutrinos. Such attempts were also made in the past
\cite{emr,rt}.  Our analysis, though similar in nature, differs
with the previous ones in the neutrino mass spectrum used and
complements them by including a relevant piece of the Lagrangian
overlooked in the existing literature.

There are two issues that one needs to address. One is the origin
of the neutrino mass and the other is the interaction that drives
its decay.  We have considered a four-neutrino (three standard
and one sterile) framework which permits mass-splitting (or for
that matter even the mass of the decaying neutrino) as large as
order 100 keV or so, remaining consistent with the oscillation
data and the recent result of the tritium beta-decay experiment.
The authors of ref.~\cite{emr} have restricted themselves to a
three-neutrino framework and considered neutrino masses not more
than order 1 eV. In fact, as we argued in section 2, a
three-neutrino scenario, given the present constraints from
oscillation and tritium beta-decay, cannot permit a neutrino to
have a mass more than order 1 eV. In this situation it is not
possible to improve on the existing constraints on the $\Rsl$
couplings. Obviously the four-neutrino model provides some
breathing space allowing at least one neutrino to have a
significantly larger mass decaying (with more phase space) to the
lighter species. The resulting constraints, depending on the mass
of the decaying neutrino, are indeed tighter. In the analysis of
refs.~\cite{emr,rt}, the neutrino Majorana masses are generated
only by $\Rsl$ interactions and the same couplings eventually
lead to their decay. In our case, on the contrary, the neutrinos
receive masses (either Dirac- or Majorana-type) from some other
origin. In fact, we have put these masses in by hand, holding the
$\Rsl$ interactions accountable only for their decay. If
instead one assumes that the $\Rsl$ interactions responsible
for neutrino decay also generate the mass of the decaying
neutrino then the limits we obtained will be altered. This can be
seen as follows. The combinations $\lambda'_{ijk}
\lambda'_{i'kj}$ that have been constrained by our present
analysis also contribute to the off-diagonal ($ii'$) terms of the
left-handed (neutrino) Majorana mass matrix, while the
combinations $\lambda'_{ijk} \lambda'_{ikj}$ contribute to the
diagonal ($ii$) term. Assuming that $m_{\nu_H}$ in
Eq.~(\ref{rl2}) is generated in this manner and, to make a simple
estimate, choosing the off-diagonal ($ii'$) element of the matrix
about a tenth of this diagonal term, we can derive an upper bound
on $\lambda'_{ijk} \lambda'_{i'kj}$ demanding consistency with
the cosmological constraint in Eq.~({\ref{eq:t0}). In this way,
the strongest limit that we obtain is $\lambda'_{323}
\lambda'_{232} \leq 10^{-5}$.
This indicates that if $\Rsl$ interactions alone are
responsible both for the generation of neutrino mass as well as
for its decay, then cosmological constraints imply that the mass
so generated would be comparable to 45 eV.

The other point we would like to stress upon is that from the
$\lambda'_{ijk} L_i Q_j D^c_k$ term in the superpotential we get two
kinds of squark-mediated processes, one involving the scalar from
$Q_j$ and the other the one from $D^c_k$. They correspond to two
different pieces in the Lagrangian. For the radiative decay there
is an exact cancellation between these two contributions for
$j=k$. This point has been overlooked in \cite{emr,rt}. It is in
this situation that the $Z$-mediated penguins provide the only
decay mode of the heavy neutrino. Similar arguments can be
advanced for the $\lambda$ couplings.

It should be noted that the product couplings which drive the
$\nu_{i'} \rightarrow \nu_i$ decay are also responsible for a similar
process involving their SU(2) partners, namely, the $l_{i'}
\rightarrow l_i$ decay (where $l$ is a charged lepton). We have
checked that the constraints we have derived on the product couplings
in the context of neutrino decays are consistent with the bounds on
the same combinations derived from the non-observation of the charged
lepton decays \cite{mutoe}.

The $\nuH \rightarrow \nu \nu \bar{\nu}$ mode which we have
discussed arises from a flavor-violating $Z$-$\nuH$-$\nu$
coupling generated by penguin diagrams involving $R$-parity
violating couplings. Such a non-diagonal $Z$ vertex can also be
produced at tree level due to (a) neutrino-neutralino mixing
arising from the misalignment between the $\mu_{i}$ and the
sneutrino vacuum expectation values, or (b) through mixing
between active and sterile neutrinos. The magnitude of these
mixings depend on details of the model which are beyond the scope
of the present analysis. However, one can readily see that the
lifetime of a 45 eV neutrino decaying through such an
interaction is approximately $\tau_\mu (m_\mu/m_{\nu_{H}})^5
\sin^2\phi$ (here $\sin\phi$ parametrizes the suppression at the
off-diagonal $Z$ vertex), which is much larger than the age of the
universe. As a result, such decays can never compete with the
radiative decay modes which yield the bounds which we have presented.

It should be noted that neutrino oscillations will not affect the
results shown in Tables 1 and 2. Since the decay times that we find
are larger than the age of the universe while the typical oscillation
times, as required by solar and atmospheric data, are at most a few
minutes, there is little scope of repopulation of the $\nu_H$ states
from the daughter neutrinos through oscillation.  However, a very
large number of oscillations will have taken place before the decay so
that the decaying states will have an averaged effect, the probability
of an initial $\nu_H$ remaining so being $\cos^2\theta$, where
$\theta$ is a mixing angle. If to start with an equal number of
neutrinos of all types exist then this averaging will leave this
balance unaltered.

Before closing, a few comments on the existing constraints from
baryogenesis \cite{baryo} are in order. The requirement that GUT-scale
baryogenesis does not get washed out imposes $\lambda'' \ll 10^{-7}$
for generic indices, although it has been argued that such bounds are
model dependent and can be evaded \cite{dr_ross}. In our analysis we
have assumed all $\lambda''$ to be zero -- this also solves the proton
decay problem. In the absence of the $\lambda''$ couplings, the
$\lambda'$ or $\lambda$ couplings alone cannot wash out the initial
baryon asymmetry, unless there are $B$ violating but $(B-L)$
conserving sphaleron-induced non-perturbative interactions. Since the
latter interactions conserve $(\frac{1}{3} B - L_i)$ for each lepton
generation, the assumption that not all lepton numbers are
simultaneously violated suffices to preserve the initial baryon
asymmetry.  Keeping this mind, whenever in the present work we have
considered a product of two $\lambda'$ (or $\lambda$) thereby
violating two lepton numbers, we have made an implicit assumption that
the third lepton number is conserved.

\vskip 10pt \noindent 
\section*{Acknowledgements}
GB acknowledges hospitality of the Max-Planck Institute in Heidelberg,
and CERN Theory Division, Geneva, where some parts of the work were
done.  SR acknowledges support from the Council of Scientific and
Industrial Research, India. AR has been supported in part by the
Council of Scientific and Industrial Research and the Department of
Science and Technology, India.




\begin{thebibliography}{99}
\bibitem{t0} According to the Big Bang model of cosmology $t_0$ is
estimated to be $11.5+1\pm1.5$~Gyr \cite{pdg}.
\end{thebibliography}
\end{document}